\input{aipcheck}

\documentclass[
    ,final            % use final for the camera ready runs
  ]
  {aipproc}

\layoutstyle{6x9}

\begin{document}

\title{Double Neutron Stars: Evidence For Two Different Neutron-Star Formation Mechanisms}

\classification{04.40.Dg, 97.60.Jd, 97.60.Gb}

%%<Replace this text with PACS numbers;choose from this list:
%%          \ttext{http://www.aip..org/pacs/index.html}>

\keywords {stars: neutron --- stars: magnetic fields --- stars: relativistic 
--- pulsars: general}

\author{E.P.J. van den Heuvel}{address={Astronomical Institute ``Anton Pannekoek'' 
and Center for High Energy Astrophysics, University of Amsterdam, The Netherlands, 
and Kavli Institute for Theoretical Physics, University of California, Santa Barbara, 
USA},,email={edvdh@science.uva.nl}}

\begin{abstract}
Six of the eight double neutron stars known in the Galactic disk have
low orbital eccentricities ($< 0.27$) indicating that their
second-born neutron stars received only very small velocity kicks at
birth. This is similar to the case of the B-emission X-ray binaries,
where a sizable fraction of the neutron stars received hardly any
velocity kick at birth (Pfahl et al.\ 2002). The masses of the
second-born neutron stars in five of the six low-eccentricity double
neutron stars are remarkably low (between 1.18 and
1.30M$_{\odot}$). It is argued that these low-mass, low-kick neutron
stars were formed by the electron-capture collapse of the degenerate
O-Ne-Mg cores of helium stars less massive than about 3.5M$_{\odot}$,
whereas the higher-mass, higher kick-velocity neutron stars were
formed by the collapses of the iron cores of higher initial mass. The
absence of low-velocity single young radio pulsars (Hobbs et al. 2005)
is consistent with the model proposed by Podsiadlowski et al.\ (2004),
in which the electron-capture collapse of degenerate O-Ne-Mg cores can
only occur in binary systems, and not in single stars.
\end{abstract}

\maketitle{}

\section{The birth kick velocities of neutron stars}

Pfahl et al. (2002) discovered the existence of a separate class of
B-emission X-ray binaries (abbreviated here as Be/X-ray binaries) with
wide orbits of low eccentricity ($< 0.25$). The systems in this class
tend to have relatively low X-ray luminosities ($< 10^{34}$ ergs/s). A
well-known example is X-Per, in which the neutron star has an almost
circular orbit with a period of 250 days. About half of all Be/X-ray
binaries with known orbits appear to belong to this class and the
relatively low X-ray luminosities of these sources imply that these
systems are on average considerably nearer to us than the
high-eccentricity Be/X-ray binaries (which during outbursts can reach
a luminosity of $10^{38}$ ergs/s). Therefore, as Pfahl et al. (2002)
pointed out, the systems in the low-eccentricity class probably form
the bulk of the Be/X-ray binary population, since the known numbers of
sources in both classes are about the same.  These authors pointed out
that the neutron stars in the low-eccentricity systems cannot have
received a kick velocity at their birth exceeding 50~km/s. Until the
discovery of this class of X-ray binaries it was generally thought
that all neutron stars receive a high kick velocity at their birth, of
order at least a few hundred km/s (see e.g.: Lyne and Lorimer 1994;
Hansen and Phinney 1997, Hobbs et al.2005). Often a Maxwellian
distribution is used to represent the observed distribution of pulsar
velocities, and the characteristic velocity of these Maxwellians is
typically around 300 -- 400 km/s (Hansen and Phinney 1997).

A recent very detailed study by Hobbs et al. (2005) of the accurately
determined proper motions of 233 radio pulsars shows that there is no
room for a separate population of low-velocity single
pulsars. Particularly, these authors found that the velocity
distribution of young pulsars (age $<$ 3 million years) is very well
represented by a single Maxwellian with a characteristic velocity of
about 400 km/s, and there is no evidence for a bimodal velocity
distribution as had been argued by Cordes and Chernoff (1998).

On the other hand, Pfahl et al. (2002) showed, by means of population
synthesis calculations that include the evolution of binaries and the
presence of birth kicks imparted to neutron stars, that with the
assumption of only one Maxwellian with a high characteristic velocity
(several hundred km/s) one can reproduce the high-eccentricity
population of the Be/X-ray binaries, but one totally fails to
reproduce the presence of a large population of systems with low
eccentricities. They convincingly showed that the only way in which
both the observed high-$e$ and the low-$e$ populations of the Be/X-ray
binaries can be reproduced is: by assuming that there are two distinct
populations of neutron stars: one population that receives hardly any
kick velocity at birth ($v_{k} < 50$~km/s) and another which receives
the ``canonical'' high velocity kick of order several hundreds of km/s
at birth.

\section{Double neutron stars and the low kick velocity neutron star population}

At present 9 double neutron stars are known, 8 of them in the galactic disk 
and one in a globular cluster (see Stairs 2004, Lorimer et al. 2006). 
The eight systems in the galactic disk are listed in table 1.
As the table shows, the double neutron stars tend to have very narrow orbits. 
They are the later evolutionary products of wide high-mass X-ray binary systems 
with orbital periods $>$ 100 days (van den Heuvel and Taam 1984), 
mostly B-emission X-ray binaries (for an alternative view, see Brown 1995). 
When the massive star in such a system has 
expanded to become a red giant, its envelope engulfs the neutron star, 
causing this star to spiral down into this envelope, reducing its orbital 
separation by several orders of magnitude. The large energy release due 
to friction and accretion during this spiral-in process is expected to 
cause the hydrogen-rich envelope of the giant to be expelled such that a 
very close binary remains, consisting of the helium core of the giant 
together with the neutron star (van den Heuvel and Taam 1984; Dewi and Pols 2003). 
(Depending on the orbital separation at the onset of spiral in, the helium core 
itself may already be (somewhat) evolved and possibly contain already some C and O in its core).  
[In Be/X-ray systems that started out with orbital periods $< 100$ days the neutron star 
spirals in so deeply that it most probably merges with the core of the giant, 
and so no binary will be left; e.g. see Taam 1996].
Due to the large frictional and tidal effects during spiral in the orbit of 
the system is expected to be perfectly circular. 
The helium star generates its luminosity by helium burning, 
which produces C and O, and subsequently by carbon burning, producing Ne and Mg.
 
If the helium star has a mass in the range 1.6 to ~3.5 M$_{\odot}$
(corresponding to a main-sequence progenitor in the range 8 to 11
($\pm{1}$) M$_{\odot}$, the precise limits of this mass range
depending on metallicity and on the assumed model for convective
energy transport; Sugimoto and Nomoto 1980; Miyaji et al. 1980;
Podsiadlowski et al. 2004) it will during carbon burning develop a
degenerate O-Ne-Mg core, surrounded by episodic C-and He-burning
shells (e.g. Nomoto 1984, Habets 1986). When such a degenerate core
develops, the envelope of the helium star begins to expand, causing in
a binary the onset of mass transfer by Roche-lobe overflow (Habets
1986; Dewi and Pols 2003). Roche-lobe overflow leads to the formation
of an accretion disk around the neutron star and accretion of matter
with angular momentum from this disk will cause the spin frequency of
the neutron star to increase. Therefore one expects that during the
later evolution of these helium stars of relatively low mass the
first-born neutron star in the system will be ``spun up'' to a short
spin period. This neutron star had already a long history of
accretion: first when it was in a wide binary with an early-type
(presumably Be) companion; subsequently during the spiral-in phase
into the envelope of its companion and now as companion of a
Roche-lobe overflowing helium star. Since all binary pulsars which had
a history of mass accretion (so-called ``recycled'' pulsars;
Radhakrishnan and Srinivasan 1982) tend to have much weaker magnetic
fields than normal single pulsars, it is thought that accretion in
some way causes a weakening of the surface dipole magnetic field of
neutron stars (Taam and van den Heuvel 1986) and several theories have
been put forward to explain this accretion-induced field decay
(Bisnovatyi-Kogan and Komberg 1974; see Bhattacharya and Srinivasan
1995 for a review; Zhang 1998 and Cumming 2004). 

With a field weakened
to about $10^{10}$ Gauss (as observed in the recycled components of
the double neutron stars (see table 1), and an Eddington-limited
accretion rate of helium of $\sim 4 \times 10^{-8}$ M$_{\odot}$/yr, a neutron
star can be spun-up to a shortest possible spin period of a few tens
of milliseconds (Smarr and Blandford 1976, Srinivasan and van den
Heuvel 1982).  When the helium star finally explodes as a supernova,
the second neutron star in the system is born.  This is a newborn
neutron star without a history of accretion and is therefore expected
to resemble the ``normal'' strong-magnetic field single radio pulsars
(Srinivasan and van den Heuvel 1982), which have typical surface
dipole magnetic fields strengths of $10^{12}$ -- $10^{13}$ Gauss. This
theoretical expectation has been beautifully confirmed by the
discovery of the double pulsar systems PSRJ0737-3039AB, which consists
of a recycled pulsar (star A) with a very rapid spin (P = 23 ms) and a
weak magnetic field ($7 \times 10^{9}$ G) and a normal strong-magnetic-field
($1.2 \times 10^{12}$ G) pulsar (star B) with a ``normal'' pulse period of
2.8 sec (Burgay et al. 2003, Lyne et al. 2004; see table 1). The
explosive mass loss in the second supernova has made the orbit
eccentric and since the two neutron stars are basically point masses,
tidal effects in double neutron star systems will be negligible and
there will be no tidal circularization of the orbit. (On timescales of
tens of millions of years the orbits may circularize by a few
percent due to the emission of gravitational waves in the
shortest-period system of PSRJ0737-3039, but in all the other double
neutron stars this is a negligible effect, except in the final stages
of spiraling together; see e.g. Shapiro and Teukolsky 1983).  

In case
of spherically symmetric mass ejection in the supernova there is a
simple relation between the orbital eccentricity and the amount of
mass $\Delta M_{sn}$ ejected in the supernova: 
\begin{equation}
    e = \Delta M_{sn}/(M_{ns1} + M_{ns2})
\end{equation}
where $M_{ns1}$ and $M_{ns2}$ are the masses of the first- and the second-born 
neutron stars.
The ``conventional'' kick velocities of neutron stars of about 400
km/s (Hobbs et al.2005) are quite similar to the orbital velocities of
the neutron stars in close double neutron stars such as the
Hulse-Taylor binary pulsar PSRB1913+16 (P$_{\rm{orb}}$ = 7.75
hours). Therefore, a kick velocity of this order produces a major
disturbance of the orbit and -- unless it is imparted in a very
specific direction -- will in general impart a large eccentricity to
the orbit, of order 0.5 or more. The Hulse-Taylor binary pulsar has a
large eccentricity $e$ = 0.617 and the same is true for the system
PSRJ1811-1736 ($e$ = 0.828), which indeed might be due to such large
kick velocities. However, as table 1 shows, very surprisingly all of
the other 6 double neutron stars in the galactic disk have very small
orbital eccentricities, in the range 0.088 to 0.27. Such
eccentricities are the ones which one expects from the pure sudden
mass loss effects in the supernova explosion, given by equation (1),
but not in case a randomly directed kick velocity of order 400 km/s is
imparted to the second-born neutron star at birth. [In particular, the
small orbital eccentricities of the two relatively wide double neutron
stars PSRJ1518+4909 and PSRJ1829+2456 are impossible to reconcile with
high kick velocities].  

Furthermore, Dewi et al.\ (2005) and van den
Heuvel (2005) have pointed out that the relation between spin period of
the recycled neutron star and orbital eccentricity observed in double
neutron star systems (Faulkner et al.\ 2005) can only be understood if
the second-born neutron stars in these systems received a negligible
velocity kick in their birth events. Interestingly, also the
Hulse-Taylor binary pulsar PSRB1913+16 and PSRJ1811-1736 fit this
relation, which suggests that also their high orbital eccentricities
were purely due to the effects of the sudden mass loss in the second
supernova. And indeed, since their first-born neutron stars are quite
strongly recycled, they must have had a quite extended episode of disk
accretion. This implies an extended episode of stable Roche-lobe
overflow from the helium star progenitor of the second-born neutron
star. And this in turn suggests that these helium stars had a
degenerate O-Ne-Mg core, as only the development of such cores causes
the envelopes of helium stars to expand.

It thus appears that the second-born neutron stars in these 6
low-eccentricity systems belong to the same ``kick-less'' class as the
neutron stars in the low-eccentricity class of Be/X-ray binaries (van
den Heuvel 2004, 2005, 2006).  The same holds for the young
strong-magnetic-field pulsar in the eccentric radio-pulsar binary
PSRJ1145-6545 which has a massive white dwarf as a companion (Kaspi et
al.\ 2000; Bailes et al.\ 2003; Bailes 2005). The orbital eccentricity of
0.172 of this binary shows that the neutron star was the last-born
object in the system (Portegies Zwart and Yungelson 1999, Tauris and Sennels 2000; formation of a white dwarf cannot introduce an orbital eccentricity). The low value of its
eccentricity would be hard to understand if the neutron star received
the canonical 400 km/s kick at its birth.

\begin{table}
\begin{tabular}{ccccccccc}
\hline
       &       &                &     &                &                &                &                &      \\ 
Pulsar & Spin  & P$_{\rm orb}$  &     & Compan.        & Pulsar         &  Sum of        &  B$_{\rm s}$   &      \\
 Name  & Per.  &                & $e$ &  Mass          &  Mass          &  masses        &                & Ref  \\
       &  (ms) &  (d)           &     &  (M$_{\odot}$) &  (M$_{\odot}$) &  (M$_{\odot}$) &  ($10^{10}$ G) &      \\
       &       &                &     &                &                &                &                &      \\
\hline
J0737- & 22.7  & 0.10   & 0.088   & 1.250(5)   & 1.337(5)   & 2.588(3)  & 0.7        & (1)   \\
3039A  &       &        &         &            &            &           &            &       \\
\hline
J0737- & 2770  & 0.10   & 0.088   & 1.337(5)   & 1.250(5)   & 2.588(3)  & $1.2 \times 10^{2}$ & (1)   \\
3039B  &       &        &         &            &            &           &            &       \\
\hline
J1518+ &       &        &         & 1.05       & 1.56       &           &            &       \\
4904   & 40.9  & 8.63   & 0.249   & (+0.45)    & (+0.13)    & 2.62(7)   & 0.1        & (2)   \\
       &       &        &         & (-0.11)    & (-0.45)    &           &            &       \\
\hline
B1534+ &  37.9 & 0.42   & 0.274   & 1.3452(10) & 1.3332(10) & 2.678(1)  & 1          & (3)   \\
12     &       &        &         &            &            &           &            &       \\
\hline
J1756- &  28.5 & 0.32   & 0.18    & 1.18(3)    & 1.40(3)    & 2.574(3)  & 0.54       & (4)   \\
2251   &       &        &         &            &            &           &            &       \\
\hline
J1811- &       &        &         & 1.11       & 1.62       &           &            &       \\
1736   &  104  & 18.8   & 0.828   & (+0.53)    & (+0.22)    & 2.60(10)  & 1.3        & (3)   \\
       &       &        &         & (-0.15)    & (-0.55)    &           &            &       \\
\hline
J1829+ &       &        &         & 1.27       & 1.30       &           &            &       \\
2456   &  41.0 &  1.18  & 0.139   & (+0.11)    & (+0.05)    & 2.53(10)  & $\sim 1$   & (5)   \\
       &       &        &         & (-0.07)    & (-0.05)    &           &            &       \\
\hline
J1906+ & 144.1 &  0.165 & 0.085   & ---        & ---        & 2.61(2)   & $1.7 \times 10^{2}$& (7)   \\
0746   &       &        &         &            &            &           &            &       \\
\hline
B1913+ &  59   &  0.33  & 0.617   & 1.3873(3)  & 1.4408(3)  & 2.8281(1) &  2         & (3)   \\
16     &       &        &         &            &            &           &            &       \\
\hline
J1145- &  394  &  0.20  & 0.172   & 1.00(2)    & 1.28(2)    &  2.288(3) &  $\sim 10^{2}$  & (6)   \\
6545   &       &        &         &            &            &           &            &       \\
\hline
\end{tabular}
\caption{Double neutron star binaries and the eccentric-orbit white-dwarf neutron 
star system J1145-6545.   %<-- this version will appear in List of Tables
References: (1) Lyne et al. (2004); (2) Nice et al. (1996); (3) Stairs (2004); 
            (4) Faulkner et al. (2005);  (5) Champion et al. (2004); (6) Bailes (2005);
            (7) Lorimer et al. (2006).}
\end{table}

\section{The masses of the second-born neutron stars in the double
neutron star systems and in PSRJ1145-6545}

In the eccentric white-dwarf/neutron-star system of PSRJ1145-6545 the
mass of the neutron star is known from the measurement of relativistic
effects to be 1.28(2) M$_{\odot}$\footnote{The number within
parentheses indicates the 95\% confidence uncertainty of the last
digit; the total mass of the system is 2.30 M$_{\odot}$ and the mass
of the white dwarf is at least one solar mass.} (Bailes 2005).  Also in two 
of the low-eccentricity double neutron stars the masses of both stars are
accurately known from measured relativistic effects (see Stairs 2004):
~\\ (i) in PSR J0737-3039 the second-born neutron star has M$_{\rm B}$
= 1.250(3) M$_{\odot}$ and the first-born one has M$_{\rm A}$ =
1.330(3) M$_{\odot}$ (Lyne et al. 2004). ~\\ (ii) in PSR J1756-2251 the
second-born neutron star has a mass of 1.18(3) M$_{\odot}$ and the
first-born one a mass of 1.40(3) M$_{\odot}$ (Faulkner et al. 2005).

In most of the other double neutron stars the masses of the stars are
not yet accurately known, but in two of the other low-eccentricity
systems the second-born neutron stars must be less massive than 1.30
M$_{\odot}$ for the following reasons. In all double neutron star
systems the relativistic parameter that can be measured most easily is
the General Relativistic rate of periastron advance, which directly
yields the sum of the masses of the two neutron stars (e.g. see Stairs
2004). In the low-eccentricity systems of PSR J1518+4904, PSR J1829+2456
and PSR J1906+0746 the resulting sum of the masses turns out to be
2.62, 2.53 and 2.61M$_{\odot}$, respectively. The individual masses of
the neutron stars in these systems are still rather poorly determined,
but in the first two of these three systems the already crudely
determined other relativistic parameters indicate that the second-born
neutron star has the lowest mass of the two (see references in van den
Heuvel 2004). As in these systems the sum of the masses is around 2.60
M$_{\odot}$, the second-born neutron stars in these two systems cannot
be more massive than 1.30 M$_{\odot}$.

Thus we find that in at least four of these six systems the
second-born neutron star has a low mass, in the range 1.18 to 1.30
M$_{\odot}$ and belongs to the low-kick category. And the same holds
for the second-born neutron star in the low-eccentricity
white-dwarf-neutron-star binary PSR J1145-6545, which has a mass of
only 1.28 M$_{\odot}$. Also in the system of PSR J1909+0746 the masses
of the neutron stars cannot differ much from 1.30 M$_{\odot}$.  We thus
see that in at least five cases a low (or no) kick velocity is
correlated with a low neutron star mass of on average around 1.25
($\pm$ 0.06) M$_{\odot}$.

A neutron star of 1.25 M$_{\odot}$ corresponds to a pre-collapse mass
of about 1.44 M$_{\odot}$, as during the collapse the gravitational
binding energy of the neutron star of about 0.20 M$_{\odot}$ (slightly
depending on the assumed equation of state of neutronized matter) is
lost in the form of neutrinos. So apparently the cores, which
collapsed to these second-born neutron stars, had a mass very close to
the Chandrasekhar mass.

\section{Formation mechanisms of neutron stars and possible resulting kicks}

It is long known (Miyaji et al. 1980, Sugimoto and Nomoto 1980) 
that there are two basically different ways in which neutron stars 
are expected to form, i.e.: ~\\
(i) In stars which originated in the main-sequence mass range between 8 and 
about 11 ($\pm{1}$) M$_{\odot}$, which in binaries produce helium stars in 
the mass range 1.6 to ~3.5 M$_{\odot}$ (Habets 1986, Dewi and Pols 2003),
the O-Ne-Mg core which forms during carbon burning becomes degenerate 
and when its mass approaches the Chandrasekhar mass, electron captures 
on Mg and Ne cause the core to collapse to a neutron star. Since these stars 
did not reach Oxygen- and Silicon burning, the baryonic mass of the neutron star, 
which forms in this way, is expected to be purely determined by the mass of 
the collapsing degenerate core, which is the Chandrasekhar mass. 
The gravitational mass of this neutron star is then the Chandrasekhar mass minus 
the gravitational binding energy of the neutron star, which is about 0.20 M$_{\odot}$. 
Thus a neutron star with a mass of about 1.24 M$_{\odot}$ is expected to result. ~\\
(ii) In stars initially more massive than 11 ($\pm 1$)M$_{\odot}$, the O-Ne-Mg core does 
not become degenerate and these cores proceed through Oxygen and Silicon burning 
to form an iron core. When the mass of this iron core exceeds a critical value 
it collapses to form a neutron star. The precise way in which here neutrino 
transport during core bounce and shock formation results in a supernova 
explosion is not yet fully understood. It appears that first the shock 
stalls and then several hundreds of milliseconds later, is revitalized. 
Some fall back of matter from the layers surrounding the proto neutron star 
is expected to occur (see Fryer 2004) such that the neutron star 
that forms may be substantially more massive than the mass of the collapsing Fe-core. 

In fact there are two expected mass regimes for the resulting neutron
stars: for stars with initial main-sequence masses in the range 11
($\pm 1$) M$_{\odot}$ to 19 M$_{\odot}$ the collapsing cores are
expected to be about 1.3 M$_{\odot}$, whereas for stars more massive
than 19 M$_{\odot}$ the collapsing iron core is expected to have a
mass $> 1.7$ M$_{\odot}$ (Timmes et al. 1996), leading to the
formation of neutron stars with (gravitational) masses $>$ 1.6
M$_{\odot}$. Taking some fall-back of matter into account, the neutron
stars formed from these types of iron cores may be expected to have
gravitational masses $>$ 1.3 M$_{\odot}$ and $>$ 1.7 M$_{\odot}$,
respectively.

The fact that the pre-collapse masses of the low-mass, low-kick neutron stars 
were very close to the Chandrasekhar limit suggests that these neutron stars 
are the result of the electron-capture collapse of the degenerate O-Ne-Mg cores 
of helium stars that originated in the mass range 1.6 to ~3.5 M$_{\odot}$ 
(initial main-sequence mass in the range 8 to 11 ($\pm 1$) M$_{\odot}$). 
Can one understand why such neutron stars would not receive 
a birth kick whereas those formed by the collapse of an iron core would?
While in the past neutron-star kicks generally were ascribed to asymmetric neutrino emission (e.g. Burrows and Hayes 1996), in recent years the ideas have shifted towards hydrodynamic instabilities during the explosion. For example, Scheck et al. (2004, 2005) found large-scale hydrodynamic instabilities to develop in the layers surrounding the proto neutron star during the explosion 
of a 15 M$_{\odot}$ star with a collapsing iron core, which imparted velocities up to 1000 km/s to the neutron star. On the other hand, for collapsing O-Ne-Mg cores, Kitaura et al. (2006) did not find large neutron-star velocities. This is ascribed to the facts that (a) here the ejecta mass in the immediate vicinity of the proto neutron star is very small, and (b) the explosion of the O-Ne-Mg core by neutrino heating occurs very fast (much faster than for iron cores, where the development of the explosion takes hundreds of milliseconds), not allowing hydrodynamic instabilities to develop.  
It thus appears that a difference in the purely hydrodynamic effects during these very different types of explosions may explain the differences in the kick velocities of the resulting neutron stars.

\section{Why are there no low-velocity single pulsars?}

Podsiadlowski et al. (2004) recently argued that {\em single} stars in
the mass range 8 to 11 ($\pm 1$) M$_{\odot}$ do not produce neutron
stars, for the following reason. These stars produce helium cores in
the mass range 1.6 to ~3.5 M$_{\odot}$, but when they ascend the
Asymptotic Giant Branch (AGB), their convective envelope during the
``dredge-up'' phase penetrates the helium layers surrounding their
degenerate O-Ne-Mg cores, and erodes these helium layers
away. Therefore the degenerate cores of these stars can no longer grow
by helium shell burning. These stars lose their envelopes due to the
heavy wind mass loss during the AGB phase, and are expected to leave
behind their degenerate O-Ne-Mg cores as white dwarfs. Only single
stars more massive than about 11 ($\pm 1$) M$_{\odot}$ will leave
neutron stars, formed in this case by iron core collapse. As argued
above, these neutron stars will be of the high-kick class, so all
single neutron stars are expected to be high-velocity objects, as is
indeed observed (Hobbs et al. 2005). On the other hand, as argued by
Podsiadlowski et al. (2004), an 8 to 11 ($\pm 1$) M$_{\odot}$ star in
an interacting binary system cannot reach the AGB, as already before
reaching that very extended phase, it will in a binary have lost its
hydrogen envelope by Roche-lobe overflow. Therefore, in binaries such
stars will leave helium stars with masses in the range 1.6 to 3.5
M$_{\odot}$, which will evolve to e-capture core collapse, which
according to our above-described model leaves a low-velocity neutron
star. One therefore expects these low-velocity neutron stars to {\it
only} be born in binary systems.

\section{Conclusions}

The combination of observations indicating that: {\it (i) among the
Be/X-ray binaries and the double neutron stars there is a substantial
group with low orbital eccentricities, indicating that their last-born
neutron stars received hardly any velocity kick at birth, (ii) the
low-kick second-born neutron stars in the double neutron star systems
have a low mass, $\sim 1.25$ M$_{\odot}$, and (iii) the absence of
low-velocity neutron stars in the young radio pulsar population} can
be consistently explained if the low-mass low-kick neutron stars
originate from the electron-capture collapse of the degenerate O-Ne-Mg
cores of stars that started out with main-sequence masses in the range
$\sim 8-11$ M$_{\odot}$, while the high-kick-velocity neutron stars
originated from the iron-core collapses of stars that started out with
masses in excess of $\sim 11$ M$_{\odot}$. Such an explanation is fully
consistent with the model proposed by Podsiadlowski et al.\ (2004)
according to which neutron star formation by electron-capture collapse
can {\it only} occur in interacting binaries and {\it not} in single
stars.

\begin{theacknowledgments}
This research was supported in part by the National Science Foundation
under Grant No. PHY99-07949.  I am grateful to the Leids
Kerkhoven-Bosscha Fonds and the Netherlands Research School for
Astronomy NOVA for providing financial support enabling me to take
part in the Cefalu Conference.
\end{theacknowledgments}

%% optional, to supply a shorter version of the title for the running head:
%%\chaptitlerunninghead{}

%%\inxx{} seen below, is an indexing command, for `silent' index
%% entries. \inx{} will print on page AND send term to .inx file.

\end{document}